\documentclass[manuscript]{aastex}

\begin{document}

\title{Color Variability of HBC 722 in the Post-Outburst Phases}

\author{
Giseon Baek$^1$,
Soojong Pak$^1$,
Joel D. Green$^2$,
Stefano Meschiari$^2$,
Jeong-Eun Lee$^1$,
Yiseul~Jeon$^3$,
Changsu~Choi$^3$,
Myungshin~Im$^3$,
Hyun-Il~Sung$^4$, and
Won-Kee~Park$^4$
}

\affil{$^1$School of Space Research, Kyung Hee University, 1732 Deogyeong-daero, Giheung-gu, Yongin-si, Gyeonggi-do 446-701, Korea}
\affil{$^2$Department of Astronomy, University of Texas at Austin, Austin, TX 78712, USA}
\affil{$^3$CEOU/Department of Physics \& Astronomy, Seoul National University, 599 Gwanak-ro, Gwanak-gu, Seoul 151-742, Korea}
\affil{$^4$Korea Astronomy and Space Science Institute, Daejeon 305-348, Korea}

\begin{abstract}
We carried out photometric observations for HBC 722 in SDSS $r$, $i$ and $z$ bands from 2011 April to 2013 May with a Camera for Quasars in Early uNiverse attached to the 2.1m Otto Struve telescope at McDonald Observatory. The post-outburst phenomena were classified into five phases according to not only brightness but also color variations, which might be caused by physical changes in the emitting regions of optical and near-infrared bands. 
A series of spectral energy distribution (SED) is presented to support color variations and track the time evolution of SED in optical/near-infrared bands after the outburst. 
Given two years of data, possible periodicities of $r$, $i$ and $z$ bands were checked. We found out three families of signals around $\sim$6, $\sim$10 and $\sim$1 days in three bands, which is broadly consistent with \citet{green2013}.
We also examined short term variability (intra-day and day scales) to search for evidences of flickering by using the micro-variability method. We found clear signs of day scale variability and weak indications of intra-day scale fluctuations, which implies that the flickering event occurs in HBC 722 after outburst.
\end{abstract}


\section{Introduction}
FU Orionis type objects (FUors) are a group of pre-main sequence objects showing a long-lived outburst in optical bands \citep{hk1996}. A prototype of this group, FU Orionis, flared up by 6 magnitude in $B$ band for a few months in 1936. Since then, FU Orionis has stayed bright for $\sim$80 years and dimmed only 0.015 magnitude per year \citep{kenyon2000}.

A dozen FUors have been discovered with analogous photometric features. They are located in an active star forming region and show a rapid increase of optical brightness (3--5 mag) within a few months. A ring-shaped asymmetric reflection nebula appears after outburst \citep{goodrich1987}. They are expected to have decaying time of 10--100 years \citep{bl1994} although there are small differences in each source \citep{hk1996}. Another dozen of FUor-like objects have been found via spectral diagnosis. They have a F--G supergiant spectum in optical while a K--M giant-supergiant spectrum in near-infrared wavelength region. In addition they show double-peaked line profiles and P Cygni profiles in H$\alpha$, and infrared excess in spectral energy distribution (SED) \citep{wein1991,lee2011}.

FU Orionis type outburst phenomena have been interpreted as a sudden increase in the accretion rate by a factor of 100--1000 in comparison to that in the quiescent state. Throughout a whole outburst event, $\sim$0.01 M$_{\sun}$ of disk material is supplied to the central star \citep{bl1994,hk1996}.
In this picture, material reaching to the innermost part of the circumstellar disk is dumped into the central star. This process in FUors might give rise to detectable sign of accretion and material dumping at inner edge of the disk, often dealt with ``flickering". It can also cause inhomogeneities such as cool or hot spots on stellar surface or disk instability, which are believed to be the reason of periodic or sporadic variabilities in hour -- day timescales.

HBC 722 (also known as LkH$\alpha$ 188 G4, PTF 10qpf and V2493 Cyg) is located in the dark cloud region, named ``Gulf of Mexico", in the southern part of the North America/Pelican Nebula Complex at a distance of 520 pc \citep{laugalys2011}. HBC 722 is the second FUor with well-characterized pre-outburst optical spectrum. Before the outburst, the source had characteristics of classical T Tauri star with a mass of $\sim$ 0.5 M$_{\sun}$, bolometric luminosity of 0.85 L$_{\sun}$, visual extinction of 3.4 magnitude, and a small amount of variability generally seen in Class II young stellar object (YSO) \citep{semkov2010,miller2011,kospal2011}. Its prominent H$\alpha$ features with equivalent width of 100 nm imply that the accretion activity was high even in the quiescent state \citep{ck1979}.

In 2010 July, HBC 722 produced a large amplitude optical outburst over a few months ($\Delta$$V$= 4.7 mag), and was classified as FUors \citep{semkov2010,miller2011}. After the peak of 2010 September, it got darkened by about 1.5 magnitude in $V$ for 6 months unlike FUors. \citet{kospal2011} suggested that, with this rate of decline, HBC 722 could return to the quiescent state in a year. They claimed that it was necessary to reconsider the classification of the source as FUors, or as another category of flaring YSO with a shorter outburst duration period, like EXors. They also calculated some properties of HBC 722 after the outburst. They derived the accretion rate of 10$^{-6}~$M$_{\sun} $yr$^{-1}$ and a bolometric luminosity of 8--12 L$_{\sun}$ assuming the mass of 0.5 M$_{\sun}$ and radius of 3 R$_{\sun}$. During the outburst, the luminosity rose roughly by a factor of 10 compared to the quiescence luminosity (0.85 L$_{\sun}$), which is somewhat smaller than classical FUors whose luminosity rises by a factor of 10--100, although the source often show typical spectra of FUors \citep{audard2014}.

Since then, however, HBC 722 remained in a relatively constant status with small fluctuations, maintaining the level brighter than that in the quiescent state by 3.3 magnitude (V) for a few months and then started to re-brighten \citep{semkov2012a,green2013}. According to recent study, the source has gradually regained its brightness about 1.5 magnitude in $V$ band over two years. Thus, HBC 722 can be classified as FUors \citep{audard2014}. Meanwhile, the color also has slightly changed during the re-brightening state. It became bluer by 0.2 magnitude in $R$-$I$ color compared to the constant status \citep{semkov2014}.

In this paper, we present the results of high cadence photometric observations from 2011 April to 2013 May. This observation includes the re-brightening state in optical/near-infrared wavelengths to track the re-increase of the accretion rate of HBC 722. We trace variabilities of HBC 722 in multiple timescales (year, day and intra-day) and try to find evidence of flickering. In section 2, the details of observation strategies and data reduction techniques are described. We analyze behaviors of HBC 722 by using light curves, color curves, color-magnitude diagram and color-color diagram in section 3. In section 4, we discuss the time evolution of SEDs after outburst and diagnosis of flickering with day variability and intra-day variability (IDV) check. Finally, we summarize our conclusions in section 5. In this study, we use the AB magnitude system. Also, we deal with `long term' as year- or longer timescales and `short term' as day or intra-day timescales.

\section{Observations and Data reduction}
We carried out photometric observations of HBC 722 using Camera for QUasars in EArly uNiverse (CQUEAN) attached to the 2.1 m Otto Struve telescope at the McDonald Observatory \citep{kim2011,park2012}. Using a 1024$\times$1024 pixel deep-depletion CCD chip, CQUEAN has a 4.\arcmin7$\times$4.\arcmin7 field of view with a custom-made focal reducer \citep{lim2013}. We obtained minute scale cadence images in $r$, $i$ and $z$ bands in SDSS filter system \citep{fj1996} for 60 nights from 2011 April to 2013 May. To see the short scale behavior of HBC 722, we performed continuous time series monitoring observations for 2--8 hours in 18 nights. The observation field is shown in \citet{green2013}. The log of observations are listed in Table \ref{log}.

The images were reduced with the IRAF\footnote{IRAF is distributed by the National Optical Astronomy Observatory, which is operated by the Association of Universities for Research in Astronomy, Inc., under cooperative agreement with the National Science Foundation.}$\slash$CCDRED packages. Since most of the images were exposed for less than 30 seconds, dark subtraction was unnecessary except for the 2011 December $r$ band images. Aperture photometry was conducted by using Source Extractor \citep{ba1996}. We set the aperture size with 3 times of FWHM of seeing of each night. Errors include Poisson errors, sky background fluctuations and subtraction errors for differential photometry. For the nightly averaged points, errors are denoted by standard deviation of comparison star in a night. Typical averaged error value is $\leq$ 0.01 magnitude.

Since HBC 722 is located in an active star forming region, most of the surrounding objects in our HBC 722 field are likely to be YSOs, which could show small amplitude variabilities \citep{semkov2010,green2013}. Thus for the differential photometry, we had to carefully select comparison stars in the field. We carried out variability checks on the background objects by repetitive differential photometry paring two candidates. Finally C7 and C4 (see Table \ref{calib} for details) were chosen as a comparison star and check star among 12 selections. \citet{semkov2010} showed a comparison sequence for HBC 722 field, which consists of fifteen objects in $BVRI$ bands with care for low amplitude variability. In their work, C4 and C7 were used as photometric standards and labeled as A and C, respectively, according to their convention. We present spectral energy distributions of two objects in Figure \ref{sedcom}. We took $u$ and $g$ bands data from Sloan Digital Sky Survey (SDSS) database and $r$, $i$ and $z$ bands data from this work. We also took magnitudes from Two Micron All Sky Survey (2MASS) point source catalog \citep{2mass} for $J$, $H$ and $Ks$ bands, and Wide-field Infrared Survey Explorer (WISE) source catalog for 3.4, 4.6 $\mu$m data \citep{wise}. We converted magnitudes of 2MASS and WISE from Vega to AB magnitude system. In optical/near-infrared, both the comparison and the check star do not show any special features but have blackbody-like spectrum. We could fit SEDs assuming a single temperature blackbody and obtained approximate temperature of 3400--3500 K (spectral type M2--M3) and 3500--3600 K (spectral type M3--M4) for C4 and C7, respectively.

Flux calibration was conducted using SDSS standard stars from \citet{smith2002}, taken in a photometric night during the 2012 June observing run. We considered only the zero point and airmass terms of the standard calibration formula, since secondary and higher order terms were small enough to be ignored in the flux calibration.

\section{Results}
\subsection{Monitoring brightness and color variabilities}
We observed HBC 722 from 2011 April to 2013 May, during constant and re-brightening states of the object. The upper part of Figure \ref{overall} shows light curves for our observations in $r$, $i$ and $z$ bands and Johnson-Cousins $R$ and $I$ band literature data for comparison \citep{semkov2012a,semkov2012b,semkov2014}. We converted the magnitudes of $R$ and $I$ bands from Vega to AB magnitude system using \citet{br2007}. Our observation and archival data are fairly consistent in terms of monotonic behaviors of brightness, though there are small offsets of magnitude due to differences in filter system. From 2011 April to 2013 May, HBC 722 brightened 1.8 mag in $r$ band and 1.6 magnitude in $i$ and $z$ bands including smaller scale variations. For comparison, $R$ and $I$ bands from \citet{semkov2012a,semkov2014} got brighter by 1.73 and 1.55 magnitude in the same period. The bottom part of Figure \ref{overall} presents long term color curves. At the same time, color got bluer about 0.18 mag in $r$-$i$ and 0.1 mag in $i$-$z$. $R$-$I$ color also got bluer by 0.18 mag, and from the reddest point, it differed by 0.25 mag.

We divide outburst stages of HBC 722 until 2013 May into five phases according to its brightness and color changes (see Table \ref{phase} and Figure \ref{overall}). First, Phase 1 deals from the beginning of the outburst to the first brightness peak in 2010 September. The brightness rapidly increased and color got bluer in Phase 1. The following Phase 2 is from 2010 September to 2011 February during which the brightness got fainter and color became redder. These two phases are presented in \citet{semkov2010} and \citet{miller2011}. After dimming by 1.4 magnitude ($R$) during 6 months, the source remained relatively constant in brightness for another few months with only small bounces, but color continuously got redder. We label it as Phase 3, which describes relatively constant state in the period of outburst. On the other hand, in Phase 4, HBC 722 started to re-brighten slowly from 2011 October. It steadily recovered its luminosity and the color curves showed distinctive bluer tendency again until 2012 May. Lastly, in Phase 5, the brightness continuously increased and went over its first peak of luminosity in 2013 May observation. In contrast to Phase 4, color curves of the last phase have weak bluer tendency or close to constant with small fluctuations. According to \citet{semkov2014}, HBC 722 maintains similar brightness from the secondary peak in 2013 May.

We found hints of short term phenomena distinguished from the long term behaviors. Figure \ref{201108} shows sample data collected over two weeks in 2011 August. Upper two panels are the light and color curves of HBC 722, and lower two panels are those of comparison star, C7. We averaged the data of each night to see day scale brightness and color variabilities. Note that this period belongs to Phase 3 in the long term, which HBC 722 stayed relatively constant in brightness and became redder in color (see Figure \ref{overall}). In contrast, day scale fluctuations are seen with the other short scale decreasing tendency of 0.1 mag amplitude. Besides, color hardly changed at the same time. Figure \ref{201209} is another sample data collected in 2012 September. Figure \ref{201209} is produced in the same manner with Figure \ref{201108}. This period is on Phase 5 with respect to the long term, whose luminosity got brighter and color showed slight changes. Likewise to the previous case, there are small fluctuations in light curves in the short term. 
In summary, distinguished behaviors on different timescales imply that there could be shorter timescale mechanisms in HBC 722 system.

\subsection{Color-magnitude diagram}
Based on our $r$, $i$ and $z$ bands photometry in Phase 3, 4, and 5, we present $i$ vs. $r$-$i$ and $i$-$z$ color-magnitude diagram in Figure \ref{cm_overall_2}. In the figure, the upper and lower sequences depict $i$-$z$ and $r$-$i$ colors, respectively. Each point is obtained by averaging whole data taken in a night.
During overall observed period, both $r$-$i$ and $i$-$z$ colors have become bluer as brightness has increased. The $r$-$i$ tendency is slightly steeper than $i$-$z$'s. This is an expected behavior due to the temperature change of the source. When accretion-related variability occurs, the amount of energy from the source and emission distribution with wavelengths change. This phenomenon results in the variation of SED shape. At that time, the wavelength range shorter than emission peak would lie on the blue edge or ``Wien side" of SED, and thus this part is very sensitive to the temperature change \citep[e.g.,][]{h2008}. In the case of HBC 722, the characteristic wavelengths are located at optical/near-infrared.
Thus over the long term, the bluer tendency at optical/near-infrared colors as brightness increases is reasonable.
We also analyze the data in each phase.
Color remained constant in Phase 3 and got bluer when it entered to Phase 4. In Phase 5, $i$-$z$ color remains almost constant again but $r$-$i$ color gets still bluer.
Additionally, there are local fluctuations of a few days in the phases. In the long term, the grouped phases well follow the bluer trend, but in the short term, the order of a few days variation are fluctuating within the grouped phases, which is not in regular sequence along with the long term trend.

\subsection{Color-color diagram}
Figure \ref{cc_overall} shows $r$-$i$ vs. $i$-$z$ color-color diagram of our observation data. Points averaged over a single night are presented to see day scale behaviors. Again, data are grouped according to the phases. We hardly see color variations during Phase 3, but they moved toward bluer direction with the increase in brightness in Phase 4. It remained in similar position from Phase 5, in which the color hardly varied in $i$-$z$ but got bluer in $r$-$i$.
Both Figure \ref{cm_overall_2} and \ref{cc_overall} illustrate varying color features of HBC 722 in regard to the phases. Therefore, we argue that this color changes might be caused by altering physical properties in individual phases.

\section{Discussion}
\subsection{Periodic signals}
\citet{green2013} reported two families of day scale periodic variability in SDSS $r$ band (5.8 and 1.28 day, 44 and 16 mmag amplitude, respectively) in the re-brightening phase. The authors proposed two scenarios, assuming the 5.8 and 1.28 day periods were attributed to stellar rotation period and disk instability and vice versa. Additionally, they suggest flickering rather than a fixed asymmetry at the inner disk edge, as an alternative source of the periodicities, even though the families of periodicity continue in a full year timescale. In this Section we look for evidence of short term periodic variability in the $r$, $i$, and $z$ bands and check the trend with \citet{green2013} .

We first search for periodicities by computing a generalized Lomb-Scargle periodogram \citep{zechmeister2009} for the photometry obtained in each band. The normalized power at each frequency $\omega$ is given by
\begin{equation}
P(\omega) = N_H \frac{\chi^2_0 - \chi^2(\omega)}{2\chi^2_0} = \frac{N_H}{2} p(\omega) \, 
\end{equation}\label{chi2}
where $\chi^2(\omega)$ is computed for the best-fitting sinusoidal signal at that period, $\chi^2_0$ is the weighted mean of the observations, and $N_H$ is the number of degrees of freedom. $p = 0$ indicates no improvement over the null hypothesis (no coherent signal), while $p = 1$ is a perfect fit of the data. In order to account for the long-term trend in the brightness of HBC 722 and any drifts in the photometric baseline, we separated each observing run into an independent dataset with an adjustable offset. Given that the floating offsets can affect the strength of the signal at each periodicity, we compute the power at each periodicity by self-consistently including the offsets in the minimization of the parameters.

We show the computed periodogram in Figure \ref{periodograms}. We only scan for periodicities between 0.1 and 16 days; this range takes into account the periodicities that are effectively sampled by the data. The bottom panel of Figure \ref{periodograms} shows the periodogram of the window function, which indicates the presence of spurious periodicities related to the observational cadence, displaying the usual sharp peak at 1 day (with an alias at 0.5 days).

The periodograms in Figure \ref{periodograms}, similarly to the data in \citet{green2013}, show an abundance of strong peaks (with very low false alarm probabilities) at a number of periods. The fact that HBC 722 is an active star makes the interpretation of the data in the time domain a difficult task, since it is expected to exhibit high-frequency noise and flickering to a degree. In order to attempt to tentatively identify periodicities over two years, as opposed to the underlying noise, we also characterize signals according to the goodness of the model, as computed by a cross-validation algorithm. Cross-validation algorithms can help identify overfitting or underfitting by resampling the data. We use here the ``leave-one-out'' implementation that we divide the full dataset of $N$ observations into a \textit{training set} of $N-1$ observations and a \textit{testing set} of a single observation, rotated among all observations for each band; each training set is used to derive a new fit. The goodness of the fit at each testing data point is then used to compute the combined likelihood ($\log\mathcal{L}$). The combined likelihood is defined as 
\begin{equation}
\log\mathcal{L} = \sum_i^N \log \left(\frac{y_i - f_i(\theta_i)}{\sigma_i}\right)
\end{equation}
where $y_i$ is the i-th observation comprising the testing set, $f_i(\theta_i)$ is the corresponding prediction from the model with the set of parameters $\theta_i$ derived by fitting the training set, and $\sigma_i$ is the corresponding uncertainty. A fit that has lower $\log\mathcal{L}$ compared to an alternative fit has a higher predictive power. 

To facilitate comparison with the results of \citet{green2013}, we derive a set of potential two-signal solutions using a grid search, which starts with the periodogram computed for the single frequency search (Figure \ref{periodograms}), then computes a periodogram for the residuals and adds candidate second signals to the fit. 

Table \ref{solutions} shows the best-fitting solutions for each band. Since each band probes potentially different physical phenomena, we did not combine the fits and instead computed each solution separately for each band. For each solution, we list the period(s), normalized $\chi^2$ and likelihood computed by the cross-validation algorithm ($\log\mathcal{L}$). Firstly, we note that according to the $\log\mathcal{L}$ metric, two-signal solutions are strict, marked improvements over one-signal solutions and no-signal solutions; this suggests that the models should not be overfitting the data, or be driven by small features in the photometry. Secondly, we note that there are three families of signals -- around $\sim$6, $\sim$10 and $\sim$1 days. This is broadly consistent with \citet{green2013}, which uncovered similar families of periodicities. Unfortunately, given the noisiness of the data in the frequency domain and the potential for aliasing, it is hard in practice to disentangle the different periodicities exhibited by the data. 

Figure \ref{bestfits} shows the signal harmonics and the phased data for each band. We note there is substantial scatter both within and among the different observation runs, even after the two signals with the largest amplitude are removed. The residual scatter is likely due to high-frequency noise, in accordance to the nature of HBC 722. 
\subsection{Spectral energy distribution (SED)}
To look at the deviations of spectral emissions in the post-outburst phase, we plot the multiple SEDs for several different brightness states in Figure \ref{sed}. We take \citet{semkov2012b,semkov2014} for $B$, $V$, $R$, and $I$ bands from the outburst of 2010 to 2013 May. Data in $r$, $i$ and $z$ bands are taken from this work. For near-infrared, we take $J$, $H$ and $K_{S}$ band data from \citet{kospal2011}, \citet{anto2013} and \citet{sung2013}. All literature data are converted from Vega magnitude to AB magnitude system by using the formula of \citet{br2007}. We match the archival data to our $r$, $i$ and $z$ data to construct SEDs with closest nights. The first SED is from the first peak in 2010, on the boundary between Phase 1 and Phase 2 in our criteria. We only use literature data taken in 2010 September 19 and 20. Second one represents Phase 3 which shows relatively calm state, constructed by the data of 2011 April 28, 30 and May 2. Third one is on the boundary between Phase 3 and Phase 4, when HBC 722 started to re-brighten. We use 2011 October 30 data for all wavelengths. Fourth one is on the boundary between Phase 4 and Phase 5, when the source reached to similar brightness with the first peak brightness in 2010. We use 2012 May 20 and 27 data for this state. The last one is from Phase 5 when it brighten up more than the first peak. We use the data of 2013 April 14 May 4, 5.

To begin with, there is a main difference between the SED of re-brightening period and that of the first peak. The shape of SED for re-brightening period shows redder feature than that of first peak, even at Phase 5, when the brightness got over the first peak. Therefore comparing to the original flaring, HBC 722 showed more increase in brightness but less bluer in color at re-brightening state.

From Phase 3 to Phase 5, the shape of SED also changed slightly. The gradient becomes less steeper as HBC 722 re-brightened up, which suggests the emission from shorter wavelength becomes higher. Thus in the long term, the source got bluer. It is coherent to our color analysis in previous section again.
\citet{johnstone2013} proved the time evolution of SED in outburst with their established model. When the accretion rate increases, the peak of SED moves toward a shorter wavelength which fits for a higher temperature blackbody. Also, they predicted that the first indication of heating is luminosity rise of the source at near or shorter wavelength region than the peak of SED. Our time evolution of SED in the re-brightening state shows a good agreement with their prediction.

Assuming that the emissions entirely came from disk accretion, we calculate relative accretion rate as a function of time. The bolometric luminosities limited at optical--near infrared are obtained in each phase (the same epochs of SEDs). By taking accretion luminosity formula and properties of HBC 722 used in \citet{green2013}, accretion rates are estimated. We normalize the values for the minimum brightness epoch of Phase 3 (2011 April) and present relative accretion rate change (Figure \ref{acc}). The actual accretion rate obtained at the similar epoch to Phase 3 reported in \citet{green2013} is $\dot{M}$ = 1.31$\times$10$^{-6}$ M$_{\sun}$ yr$^{-1}$.

\subsection{Flickering}
Although only small number of outbursting YSOs have been detected, these uniquely enhanced systems provide how accretion process occurs from innermost part of circumstellar disk to central star, which play an important role for understanding the evolution of YSOs.

In accordance with \citet[][and references therein.]{kenyon2000}, ``flickering" is randomly fluctuated small amplitude (0.01--1.0 mag) variations in dynamic timescales. It could be observed in cataclysmic variables and erupting YSOs. Flickering is often thought to be a signature of disk accretion. The most widely accepted origin of flickering is a temperature region between the central star and disk, which is the vicinity of inner edge of the disk. Since a large amount of disk material falls to the stellar surface from the inner disk, flickering could be evidence for inhomogeneous accretion flow \citep[e.g.,][]{shu1994,bastien2011}.

In order to check short term behaviors of intra-day and day time scales, we quantify the potential variabilities of HBC 722 in $r$, $i$ and $z$ bands. We expect to see footprints of flickering by using a micro-variability method developed by \citet{jm1997}.

\begin{equation}
C_{1}= \frac{\sigma \left ( Object - Comp1 \right )}{\sigma \left ( Comp 1 - Comp 2 \right )} ~\ and ~\ C_{2}= \frac{\sigma \left ( Object - Comp 2 \right )}{\sigma \left ( Comp 1 - Comp 2 \right )}
\end{equation}

C$_{1}$ is calculated by standard deviation of differential photometry for an object (Object) and a comparison star (Comp1), divided by that of the comparison star (Comp1) and a check star (Comp2). C$_{2}$ is also calculated in the same manner. Finally we derived the parameter C by taking the average of C$_{1}$ and C$_{2}$ \citep[e.g.,][]{romero1999,gupta2008}. Since the calculation of parameter C includes possible variations of comparison star, valid number of variability of HBC 722 can appear only if the variation of HBC 722 is superior to that of comparisons. According to \citet{jm1997}, parameter C depends on normal distribution, so we can suggest that potential variabilities exist with 90\%, 95\% and 99\% confidence level if C values are 1.64, 1.96 and 2.57, respectively. Finally, we substitute HBC 722, C7 and C4 for Object, Comp1 and Comp2, respectively.

The results of day scale variability of HBC 722 are found in Table \ref{var_day} and Figure \ref{C}. We do not include 2011 April, 2011 December, 2012 May and 2012 November runs because each covered only a few nights including non-photometric nights. For the majority of nights, C values are over 2.57, which implies HBC 722 shows variability with a 99\% confidence level in the day scale. On the other hand, C$_{r}$ in 2011 July, C$_{r}$ and C$_{i}$ in 2011 November and C$_{i}$ in 2012 September have values between 1.96 and 2.57. According to the aforementioned rule, these have 95\% confidence for their variabilities. Lastly, C$_{z}$ in 2011 July is located between 1.64 and 1.96, which is a little less confident than the others, but we still have 90\% confidence of variability.
To sum up, HBC 722 strongly shows day scale variability in 2011 August, 2012 June and 2013 May. It displays meaningful variabilities in the other months as well. Therefore, we conclude that HBC 722 is flickering. Since there is no certain relation between the length of observation and amplitude of brightness change, it is unlikely that the duration of continuous observation affects the C values.

According to the historical efforts to find clues of flickering in the outburst stage of cataclysmic YSOs, many studies focused on finding day scale periodic and aperiodic variations. However, using the properties of low mass YSOs and their Keplerian disks, we can explore IDV, which could be originated from between the innermost part of the disk and the central star. Opportunely, our observing time at the re-brightening state of HBC 722 is related to the re-stimulated disk accretion activity. Because of our short cadence monitoring observation strategy, we can also look at the behaviors of HBC 722 in intra day scale. We quantify the IDV in the same manner with the day scale variability. The results of derived C values are tabulated in Table \ref{var_intraday}.

In analysis of micro-variability of each night, we find that C$_{r}$ and C$_{i}$ on 2011 July 5, C$_{i}$ on 2012 September 2 and 2013 May 1 have between 1.96 and 2.57, which means potential variabilities with a 95\% confidence level. Additionally, C$_{r}$ on 2011 August 24, 2011 November 4, 2012 September 6 and 2013 May 5 are all between 1.64 and 1.96, which imply 90\% of potential variability. Meanwhile, many of the C values in $z$ band, and a few of $r$ and $i$ bands are lower than 1.0. Since the micro-variability method depends on the comparison and check star, these values can be caused by the brightness differences among the object (HBC 722), comparison (C7) and check stars (C4). In this analysis, IDV is less convincing than day scale variability. Note that we suggest statistical values for variability rather than specific values.

There are several previous studies of short term variability of FUors. \citet{herbig2003} observed FU Orionis and revealed $\sim$14 days of spectroscopic periodicity in P Cygni profiles, especially in H$\alpha$, lasting more than 1.5 years. The author also discovered another 3.54 days of periodic variation arising from inner structure of photosphere.
\citet{powell2012} confirmed the periodicities found by \citet{herbig2003} and suggested that the periodic phenomena continued over 10 years. Recently \citet{siwak2013} discovered 2--9 day quasi-periodic features in FU Orionis using the Microvariability and Oscillations of STars ($MOST$) satellite. Furthermore, \citet{siwak2013} stated that it could be caused by a dump of plasma, or magneto-rotationally unstable heterogeneities in the localized accretion disk rotating at different Keplerian radii.
Meanwhile, \citet{clarke2005} reported day scale non-periodic fluctuations in another classical FUors. They detected photometric variabilities with amplitudes of 0.1 and 0.3 mag ($V$) for V1057 Cyg and V1515 Cyg, respectively, which might be caused by flickering events.

In HBC 722, day scale non-periodic variabilities are detected similar to the V1057 Cyg and V1515 Cyg cases. Because two classical FUors are brighter than HBC 722 in optical, it is plausible that HBC 722 is flickering with smaller amplitude. We found the variabilities last more than two years with diverse amplitudes.

Looking back to other studies of IDV for flaring YSOs, \citet{kenyon2000} collected photometric data of FU Orionis and detected random brightness fluctuations in a dynamic timescale of a day or less, with 0.035--0.1 mag amplitude ($V$). \citet{bastien2011} conducted rapid cadence time series photometry for V1647 Orionis in the outburst stage, which belongs to another eruptive YSO, EXors. As a result, 0.13 day (51 mmag amplitude) periodic variability was found. They believed that the periodic variability would be related to flickering by detecting `flickering noise' signs in the power spectrum of the light curves. Since HBC 722 still remains enhanced, further investigation with sufficient data will more clearly reveal IDV and properties of intra-day scale flickering.

In Figure \ref{C}, we attempt to find a relation between the long term brightness change and the short term variability. Two of the C values around 2011 August and 2012 September are particularly large, and they lie at the boundary between the phase lines (noted in Table \ref{phase}). These larger points near the boundary of phases might be caused by a transition to the next phases, as one of possible interpretation. It can be proved if additional short term variabilities are detected when the long term brightness tendency changes.
Finally, we find a tentative relation that C values get lower from $r$ to $z$, implying smaller amplitude of variation in longer wavelength. A few explanations might be possible: first, this tendency can be simply due to the brightness effect. Under assumption of no intrinsic variabilities for two objects, brighter star shows smaller brightness deviation than dimmer one. In our case, HBC 722 clearly shows intrinsic variability but this brightness effect would be added.
In other words, in $r$, $i$ and $z$ bands HBC 722 is brighter than comparison stars but the brightness difference in $z$ band is greater than that in $r$ band. Since C values that we used is obtained by relative amount of brightness deviation from that of comparisons, the values could be affected by brightness difference in each band. Therefore the C values could decrease with increasing wavelengths.
Second, intrinsic characteristics of HBC 722 could appear.
\citet{sung2013} reported that HBC 722 showed strong correlation between flux variation and fading period after outburst. They estimated the flux variations from maximum brightness in 2010 to minimum in 2011 and compared them with fading periods. As a result, both flux variation and fading timescale are larger at shorter wavelengths. The flux variation tends to decrease linearly as wavelength increases in the optical/near-infrared range. Therefore we could expect that there is an inverse proportion relation between wavelengths ($r$, $i$ and $z$ bands) and amplitudes.

\section{Conclusion}

We observed HBC 722 in SDSS $r$, $i$ and $z$ bands from 2011 April to 2013 May with CQUEAN attached to the 2.1m Otto Struve telescope at McDonald Observatory. The photometric results show that HBC 722 have re-brightened for two years and presented unprecedented high brightness. The color also have become bluer at the same period. However, the brightness and color occasionally maintained similar status rather than steadily changed, which could be related to physical processes at inner disk.
Thus we divide the post-outburst period into five phases according to brightness and color variations. We analyze color-magnitude diagram and color-color diagram to depict tendency along with the phases and possible day scale variabilities. Different shapes of optical/near-infrared emissions between at the first peak and henceforward are shown in spectral energy distribution. Additionally, multi-epoch SED shapes indicate that HBC 722 has become hotter as brightness have increased.

We conducted periodicity check for HBC 722. As shown in Figure \ref{periodograms}, HBC 722 exhibit high-frequency noise and flickering, as expected because it is in enhanced state. Thus we statistically show best-fitting solutions for each band, which were broadly converged in three families of signals around $\sim$6, $\sim$10 and $\sim$1 days. These results are well-matched with the discovered periods of \citet{green2013}.

We also investigate short term variability separated from long term variations to find indications of flickering. 
Intra-day and day scale variabilities are quantified by using micro-variability method, which takes relative amount of variability of HBC 722 to that of comparison star. 
We find clear evidences of day scale variabilities and weaker signs of IDV in $r$, $i$ and $z$ bands. Comparing these short term variabilities with long term brightness variations from Phase 3 to Phase 5, derived C values of variability tend to display larger number on nearly boundary of phases. We suggest that there could be transitions of physical processes at the innermost part of disk which attribute to the change of brightness and color behaviors.

\acknowledgements

This work was supported by the National Research Foundation of Korea (NRF) grant, No. 2008-0060544, funded by the Korea government (MSIP).
The authors thank Prof. Sang-Gak Lee and Prof. Tae Seog Yoon for valuable discussions and suggestions. This paper includes data taken at The McDonald Observatory of The University of Texas at Austin. This research has made use of the USNOFS Image and Catalogue Archive operated by the United States Naval Observatory, Flagstaff Station {http://www.nofs.navy.mil/data/fchpix/).

\clearpage

\begin{deluxetable}{cccccccc}
\tabletypesize{\scriptsize}
\rotate
\tablecolumns{8}
\tablewidth{0pc}
\tablecaption{Observing log.\label{log}}
\tablehead{
\colhead{Date\tablenotemark{a}} & \colhead{Time\tablenotemark{a}} & \colhead{JD} & &  \colhead{Exposure Time $\times$ Number of Frames\tablenotemark{b}} &  \\
\colhead{yyyy-mm-dd} & \colhead{hh:mm:dd} & \colhead{(+2450000)} & \colhead{$r$} & \colhead{$i$} & \colhead{$z$}}
\startdata
2011-04-26	&	10:51:58	&	5677.9528 	&	20$\times$24	&	20$\times$24	&	20$\times$24	\\ 		
2011-04-30	&	10:59:49	&	5681.9582 	&	15$\times$30	&	15$\times$30	&	15$\times$30	\\
2011-07-05	&	5:44:11	&	5747.7390 	&	10$\times$311	&	10$\times$311	&	10$\times$311	\\ 				
2011-07-06	&	8:23:28	&	5748.8496 	&	10$\times$5	&	10$\times$5	&	10$\times$5	\\ 		
2011-07-07	&	8:28:53	&	5749.8534 	&	10$\times$5	&	10$\times$5	&	10$\times$5	\\ 				
2011-07-08	&	8:20:31	&	5750.8476 	&	10$\times$5	&	10$\times$5	&	10$\times$5	\\				
2011-07-09	&	8:06:10	&	5751.8376 	&	10$\times$5	&	10$\times$5	&	10$\times$5	\\ 			
2011-07-11	&	7:32:59	&	5753.8146 	&	10$\times$47	&	10$\times$47	&	10$\times$47	\\ 			
2011-08-19	&	8:33:37	&	5792.8567 	&	20$\times$25	&	10$\times$25	&	10$\times$25	\\		
2011-08-21	&	6:24:41	&	5794.7671 	&	20$\times$15	&	10$\times$15	&	10$\times$15	\\		
2011-08-22	&	6:39:34	&	5795.7775 	&	20$\times$15	&	10$\times$15	&	10$\times$15	\\ 	
2011-08-23	&	4:38:42	&	5796.6935 	&	20$\times$15	&	10$\times$15	&	10$\times$15	\\		
2011-08-24	&	2:35:07	&	5797.6077 	&	20$\times$1125	&	\nodata	&	\nodata	\\
2011-08-25	&	4:37:44	&	5798.6929 	&	20$\times$15	&	10$\times$15	&	10$\times$15	 \\
2011-08-26	&	5:12:55	&	5799.7173 	&	20$\times$12	&	10$\times$12	&	10$\times$12	\\				
2011-08-27	&	5:30:30	&	5800.7295 	&	20$\times$15	&	10$\times$15	&	10$\times$15	\\ 				
2011-08-28	&	4:55:42	&	5801.7054 	&	20$\times$15	&	10$\times$15	&	10$\times$15	\\ 				
2011-08-29	&	5:17:58	&	5802.7208 	&	20$\times$15	&	10$\times$15	&	10$\times$15	\\ 				
2011-08-30	&	5:14:34	&	5803.7185 	&	20$\times$10	&	10$\times$10	&	10$\times$10	\\ 				
2011-08-31	&	3:34:55	&	5804.6493 	&	20$\times$791	&	\nodata	&	\nodata	\\ 			
2011-09-01	&	6:19:49	&	5805.7638 	&	30$\times$14	&	15$\times$2, 10$\times$12	&	10$\times$14	\\ 			
2011-10-30	&	3:12:20	&	5864.6336 	&	20$\times$6	&	10$\times$6	&	10$\times$6	\\ 		
2011-10-31	&	2:29:53	&	5865.6041 	&	20$\times$15	&	10$\times$15	&	10$\times$15	\\			
2011-11-01	&	2:12:01	&	5866.5917 	&	20$\times$15	&	10$\times$15	&	10$\times$15	\\			
2011-11-02	&	1:05:49	&	5867.5457 	&	20$\times$12, 30$\times$199	&	10$\times$211	&	10$\times$211	\\			
2011-11-04	&	1:07:04	&	5869.5466 	&	30$\times$116	&	10$\times$116	&	10$\times$116	\\ 			
2011-11-05	&	2:14:58	&	5870.5937 	&	30$\times$14	&	10$\times$14	&	10$\times$14	\\ 				
2011-11-07	&	0:47:35	&	5872.5330 	&	30$\times$15	&	10$\times$15	&	10$\times$15	\\ 				
2011-11-08	&	1:31:37	&	5873.5636 	&	30$\times$3, 60$\times$10	&	10$\times$3, 20$\times$10	&	10$\times$3, 20$\times$10	\\			
2011-11-09	&	0:48:21	&	5874.5336 	&	30$\times$179, 60$\times$27	&	10$\times$179, 20$\times$27	&	10$\times$179, 20$\times$27	\\ 				
2011-12-14	&	2:09:20	&	5909.5898 	&	90$\times$3	&	30$\times$3	&	30$\times$3	\\ 				
2011-12-16	&	1:16:41	&	5911.5533 	&	90$\times$5	&	20$\times$5	&	30$\times$5	\\ 				
2012-05-27	&	8:28:31	&	6074.8531 	&	20$\times$45, 30$\times$83	&	10$\times$128	&	10$\times$128 \\ 				
2012-05-30	&	7:24:56	&	6077.8090 	&	15$\times$300	&	4$\times$300	&	4$\times$300	\\ 				
2012-05-31	&	9:51:59	&	6078.9111 	&	10$\times$3	&	5$\times$3	&	5$\times$3	\\ 			
2012-06-26	&	7:54:30	&	6104.8295 	&	15$\times$30	&	4$\times$30	&	4$\times$30	\\			
2012-06-27	&	6:58:05	&	6105.7903 	&	15$\times$29	&	4$\times$29	&	4$\times$29	\\			
2012-06-28	&	5:09:22	&	6106.7148 	&	15$\times$222, 20$\times$177	&	4$\times$222, 6$\times$177	&	 4$\times$222, 6$\times$177	 \\ 				
2012-06-29	&	5:43:15	&	6107.7384 	&	15$\times$30	&	4$\times$30	&	4$\times$30	\\ 			
2012-06-30	&	5:29:25	&	6108.7288 	&	30$\times$30	&	8$\times$30	&	8$\times$30	\\ 			
2012-07-01	&	7:10:07	&	6109.7987 	&	30$\times$182	&	8$\times$182	&	8$\times$182	\\ 				
2012-07-02	&	6:33:15	&	6110.7731 	&	15$\times$3, 30$\times$29	&	4$\times$3, 8$\times$29	&	4$\times$3, 8$\times$29	\\
2012-09-01	&	2:58:50	&	6171.6242 	&	15$\times$27, 20$\times$ 24, 30$\times$277	&	4$\times$328&4$\times$328	\\ 			
2012-09-02	&	2:51:35	&	6172.6192 	&	30$\times$77, 10$\times$259, 15$\times$78, 20$\times$78 &4$\times$414, 6$\times$78 	&	 4$\times$336, 6$\times$156	 \\
2012-09-03	&	4:44:14	&	6173.6974 	&	20$\times$198, 30$\times$21&4 $\times$150, 5 $\times$48, 8$\times$21&5 $\times$150, 6$\times$48, 8$\times$3,10$\times$18 	\\ 	
2012-09-04	&	8:00:18	&	6174.8335 	&	15$\times$14	&	4$\times$14	&	4$\times$14 \\			
2012-09-05	&	7:14:20	&	6175.8016 	&	10$\times$38	&	4$\times$38	&	5$\times$38	\\ 			
2012-09-06	&	3:37:17	&	6176.6509 	&	30$\times$23	&	8$\times$23	&	10$\times$23	\\ 			
2012-09-07	&	2:35:53	&	6177.6083 	&	10$\times$31	&	4$\times$31	&	6$\times$31	\\ 				
2012-09-09	&	3:18:30	&	6179.6379 	&	15$\times$162, 30$\times$15	&	4$\times$162, 8$\times$15	&	6$\times$162, 12$\times$15	\\				
2012-09-10	&	2:49:43	&	6180.6179 	&	10$\times$22	&	4$\times$22	&	6$\times$22	\\				
2012-09-11	&	4:48:35	&	6181.7004 	&	10$\times$24	&	4$\times$24	&	6$\times$24	\\ 				
2012-09-12	&	6:07:20	&	6182.7551 	&	30$\times$5	&	8$\times$5	&	12$\times$5	\\ 				
2012-11-23	&	2:00:33	&	6254.5837 	&	30$\times$10	&	5$\times$10	&	5$\times$10	\\				
2012-11-24	&	1:52:13	&	6255.5779 	&	20$\times$30	&	5$\times$30	&	5$\times$30	\\				
2012-11-26	&	1:45:33	&	6257.5733 	&	20$\times$53	&	5$\times$53	&	5$\times$53	\\ 			
2013-04-30	&	9:03:50	&	6412.8777 	&	15$\times$162	&	4$\times$162	&	4$\times$162	\\ 				
2013-05-01	&	9:27:04	&	6413.8938 	&	15$\times$84, 10$\times$60	&	4$\times$144	&	4$\times$144	\\		
2013-05-04	&	8:45:14	&	6416.8648 	&	15$\times$134, 30$\times$9	&	4$\times$134, 8$\times$9	&	4$\times$134, 10$\times$9	\\				
2013-05-05	&	8:30:23	&	6417.8544 	&	15$\times$150	&	4$\times$150	&	4$\times$150	\\				
\enddata
\tablecomments{$r$ band data from 2011 April to 2012 July is mentioned in \citet{green2013}.}
\tablenotetext{a}{Date and Time in UT. Time refers when observation started.}
\tablenotetext{b}{Exposure time per frame in units of second.}
\end{deluxetable}

\clearpage

\begin{deluxetable}{ccccccccc}
\tabletypesize{\scriptsize}
\tablecolumns{9}
\tablewidth{0pc}
\tablecaption{Target list\label{calib}}
\tablehead{
\colhead{Target} & \colhead{USNO-B1.0\tablenotemark{a} ID} & \colhead{R.A. (2000)} & \colhead{Dec. (2000)} & & \colhead{AB magnitude} & & \colhead{Spectral Type} \\
& & & & \colhead{$r$} & \colhead{$i$} & \colhead{$z$}}
\startdata
HBC 722 & 1338-0391463 & 20:58:17.0 & +43:53:42.9 & 14.3 $\sim$ 12.5\tablenotemark{b} & 13.4 $\sim$ 11.8\tablenotemark{b} & 12.8 $\sim$ 11.2\tablenotemark{b} & late K--M\tablenotemark{c}\\
C4\tablenotemark{d} & 1338-0391536 & 20:58:30.0 & +43:52:23.8 & 14.31 & 13.80  & 13.50 & M2--M3\\
C7 & 1338-0391522 & 20:58:26.3 & +43:52:23.5 & 15.15 & 14.53  & 14.20 & M3--M4\\
\enddata
\tablenotetext{a}{http://www.nofs.navy.mil/data/fchpix/}
\tablenotetext{b}{Brightness changes in our observed period (2011 April -- 2013 May) are shown.}
\tablenotetext{c}{\citet{miller2011}}
\tablenotetext{d}{Coordinate and $r$ mag from \citet{green2013} are quoted.}
\end{deluxetable}

\clearpage

\begin{deluxetable}{ccccc}
\tabletypesize{\scriptsize}
\tablecolumns{5}
\tablewidth{0pc}
\tablecaption{Classification of Phases\label{phase}}
\tablehead{
 \colhead{Period}& \colhead{Date (UT)}  &\colhead{JD (+2450000)} & \colhead{Magnitude} & \colhead{Color}}
\startdata
Phase 1 & -- 2010 Sep & -- 5460 &  Brighter & Bluer \\
Phase 2 & 2010 Sep -- 2011 Feb & 5461 -- 5600 &   Fainter & Redder \\
Phase 3 & 2011 Feb -- 2011 Oct & 5601 -- 5850 &   Constant & Redder \\
Phase 4 & 2011 Oct -- 2012 May & 5851 -- 6050 &   Brighter & Bluer \\
Phase 5 & 2012 May -- & 6051 --  &  Brighter & Constant \\
\enddata
\end{deluxetable}

\clearpage

\begin{figure}
\epsscale{.80}
\plotone{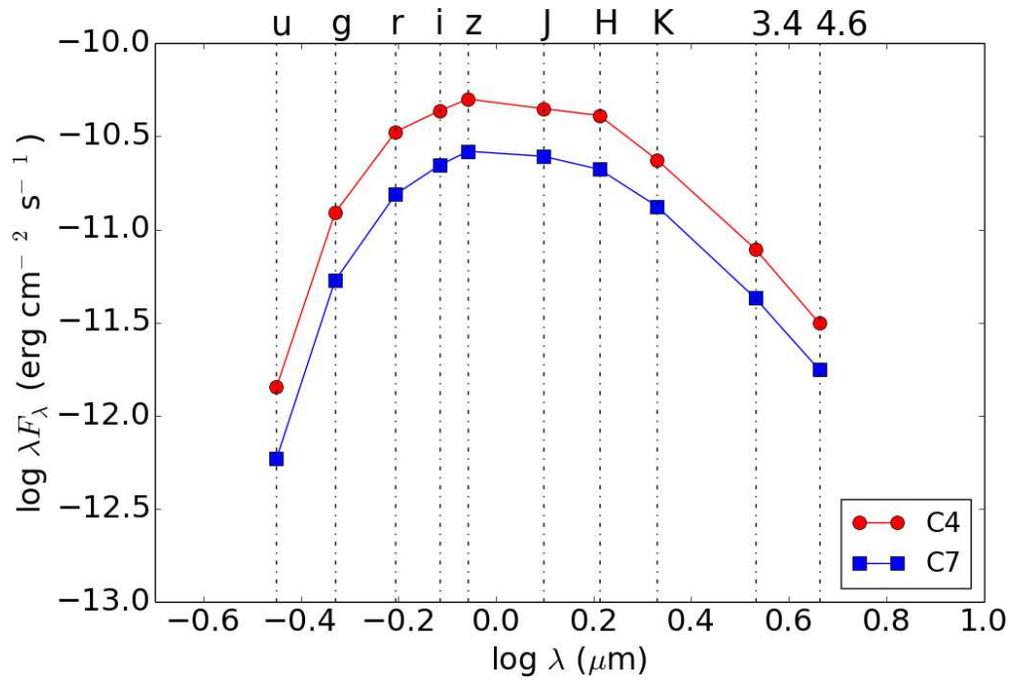}
\caption{Spectral energy distributions of comparison (C7) and check star (C4). In addition to $r$, $i$ and $z$ data from this work, archival data from SDSS, 2MASS and WISE are adopted.\label{sedcom}}
\end{figure}

\clearpage

\begin{figure}
\epsscale{.80}
\plotone{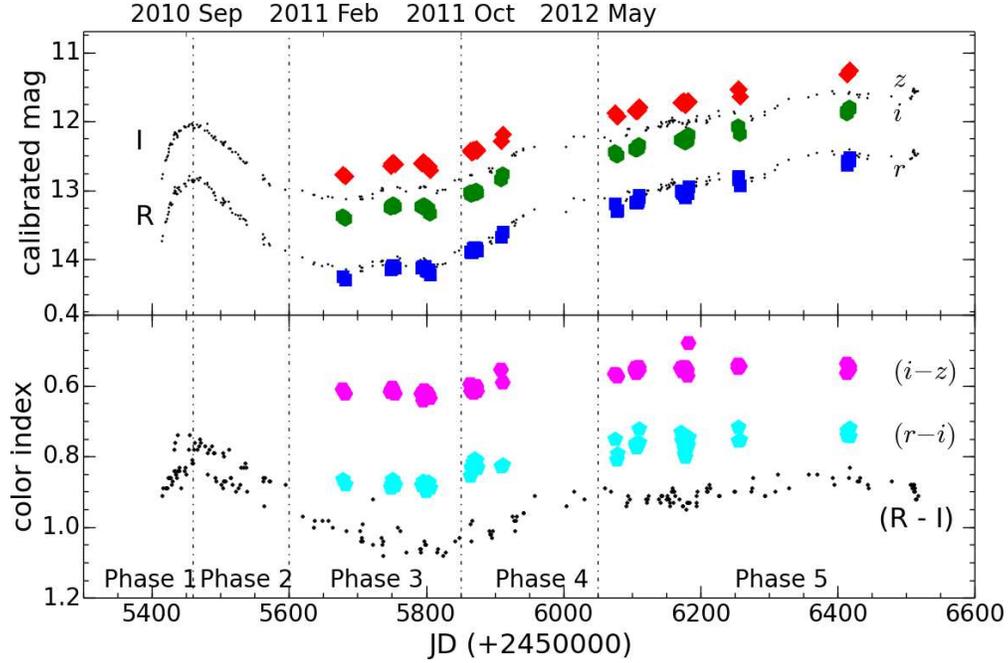}
\caption{Top: Light curves of HBC 722 collected during our observed period (2011 April -- 2013 May). The upper (red), middle (green) and the the lower (blue) symbols represent light curves of $z$, $i$ and $r$ bands, respectively. For comparison, \citet{semkov2012a,semkov2014} of $I$ and $R$ bands are also plotted in small black dots.
Bottom: Color variations. The upper (pink) and middle (cyan) dotted curves represent $i$-$z$ and $r$-$i$ color indices respectively. We also plot the $R$-$I$ color from \citet{semkov2012a,semkov2014} as references (small black dots). The small offsets between our data and references are from the different filter system.\label{overall}}
\end{figure}

\clearpage
\begin{figure}
\epsscale{.70}
\plotone{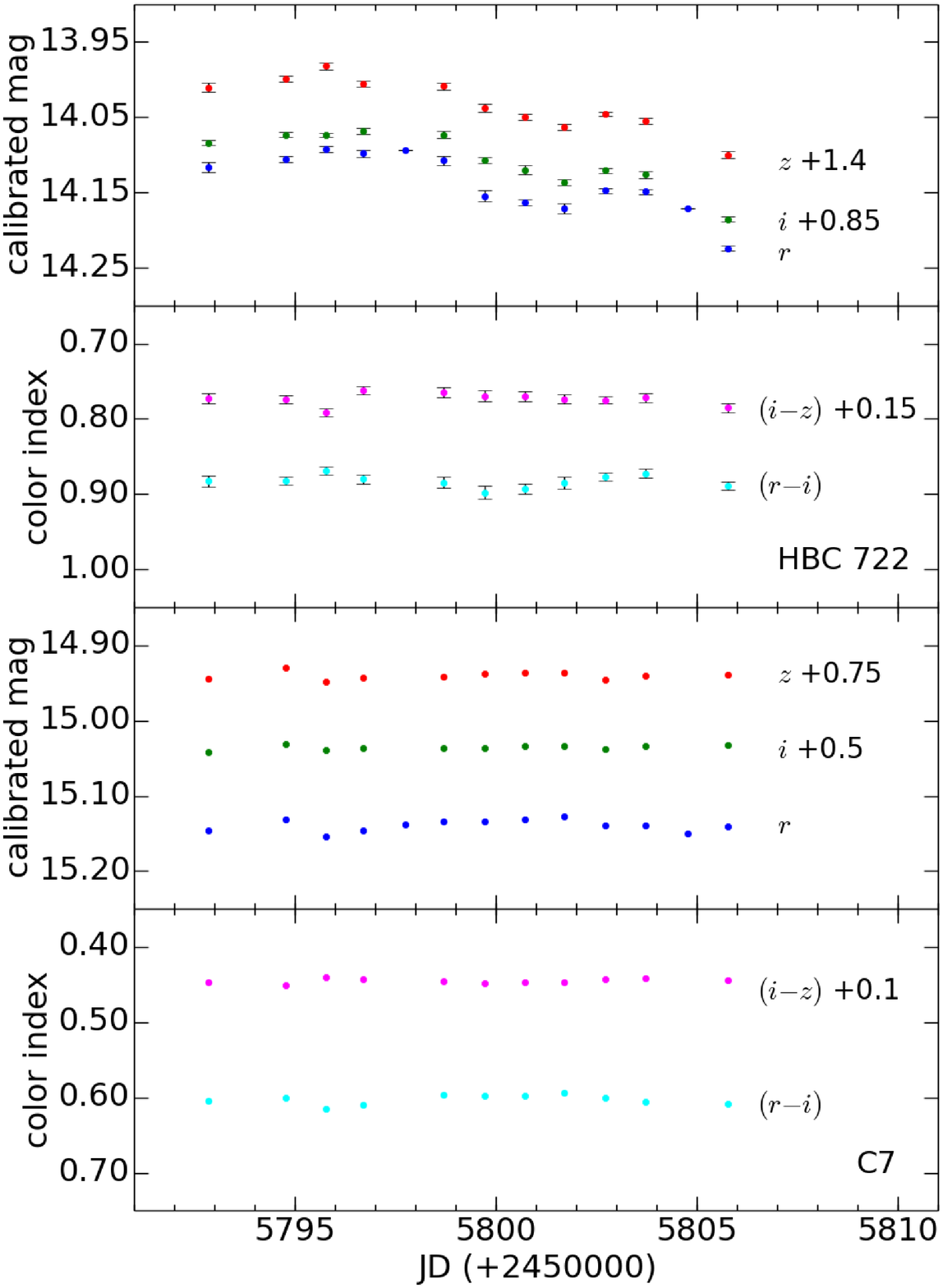}
\caption{Sample of day scale light and color curves in 2011 August (Phase 3). The top two panels are light and color curves of HBC 722 and the bottom two panels are those of comparison star C7. In the light curves, the upper (red), middle (green) and the lower (blue) dots illustrate light curves of $z$, $i$ and $r$ bands, respectively. In the color curves, upper (pink) and the lower (cyan) dots represent $i$-$z$ and $r$-$i$ color respectively. 
Data taken in a single night are averaged. For better visibility, offsets are applied. The light curves suggests the presence of day scale variabilities while color hardly changed in day scale in the same period. \label{201108}}
\end{figure}

\clearpage
\begin{figure}
\epsscale{.70}
\plotone{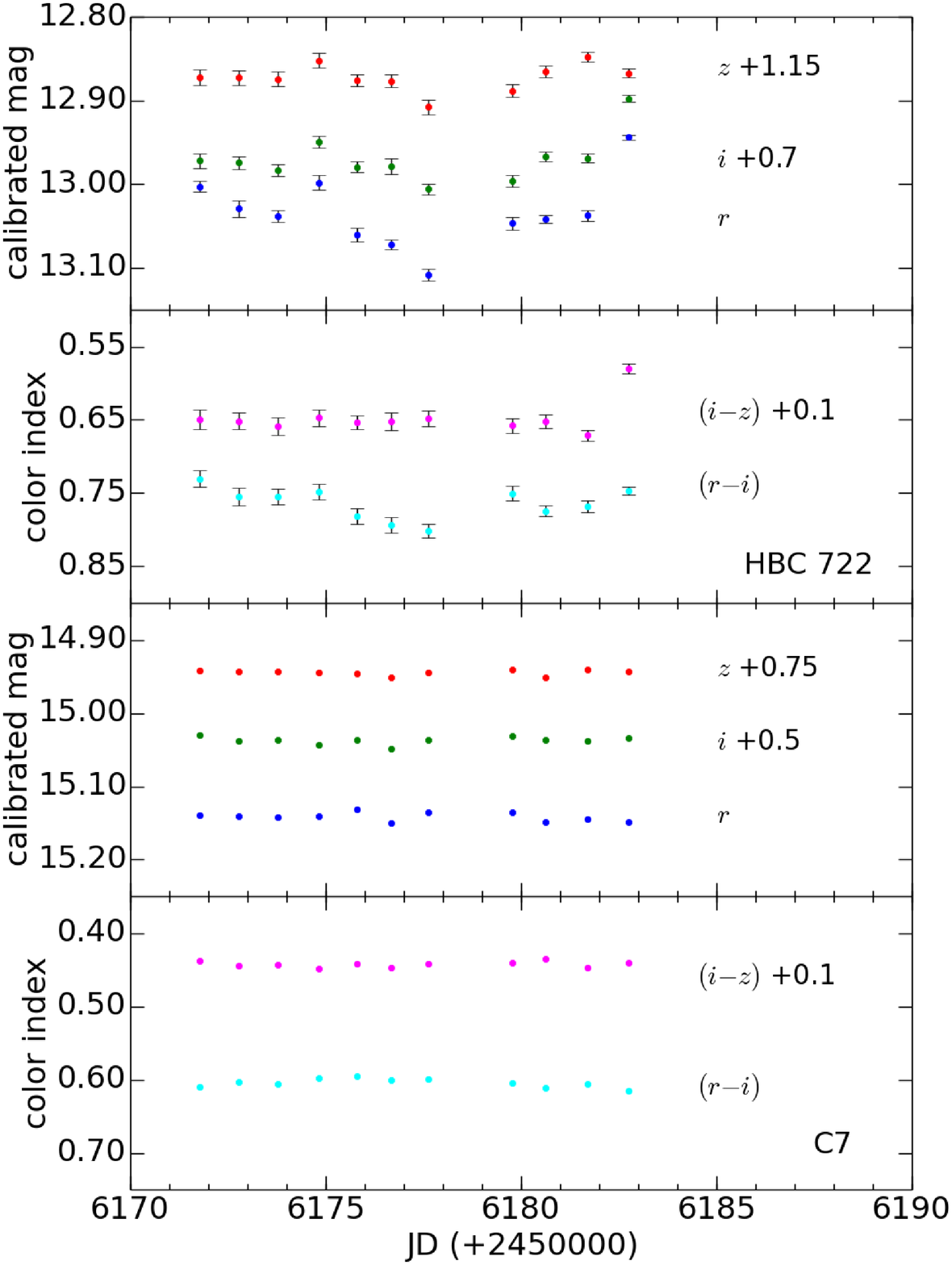}
\caption{Sample of day scale light and color curves in 2012 September (Phase 5). Description manner is the same as Figure \ref{201108}. There is a set of peculiar points on 2012 September 12 (JD $\sim$ 2456182.76). We did not discard these points based on the stable sequence of the comparison stars at the same time as seen in the third panel, though we note that the weather on the night was in non-photometric condition. \label{201209}}
\end{figure}

\clearpage
\begin{figure}
\epsscale{.80}
\plotone{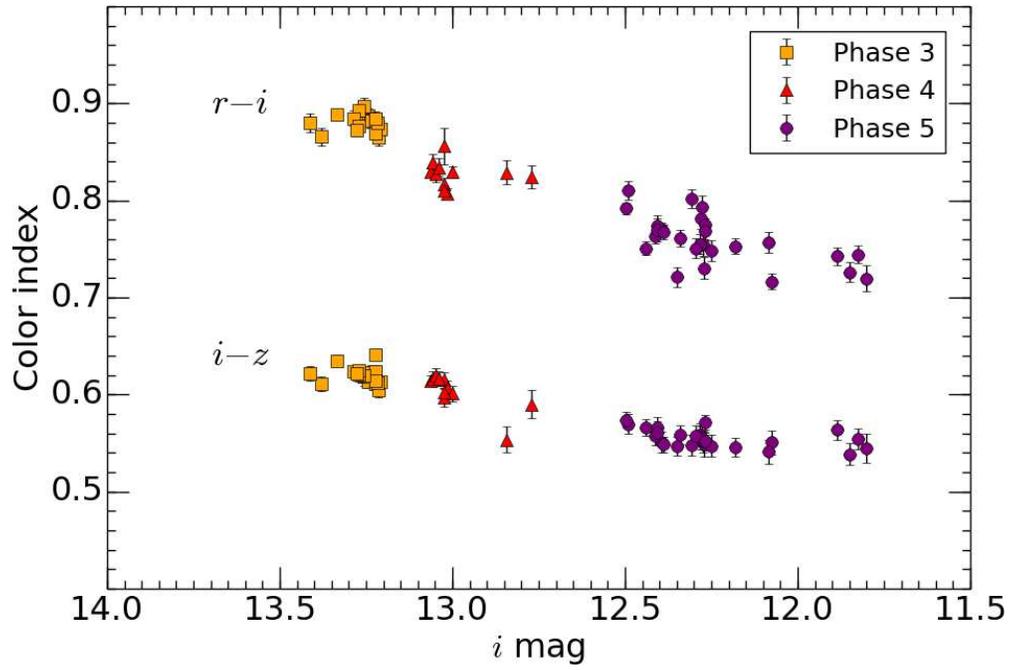}
\caption{Color-magnitude diagram during the observed period. The upper and lower layers show $i$-$z$ and $r$-$i$ color indices respectively. To look at day scale color variation, the entirety of the data taken in a night are averaged. \label{cm_overall_2}}
\end{figure}

\clearpage
\begin{figure}
\epsscale{.80}
\plotone{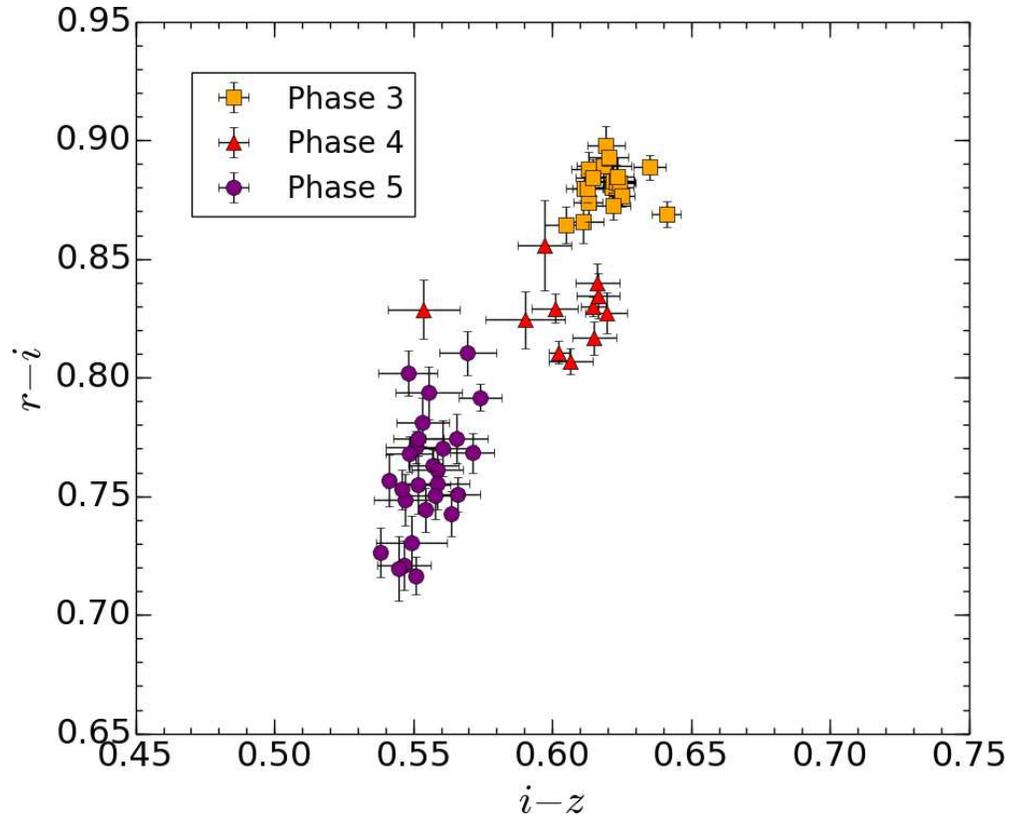}
\caption{Color-color diagram for observed period. Each dot represents the average value in a night. We can find that the color hardly moved during Phase 3 (2011 February -- 2011 August) and start to get bluer while the source brighten in Phase 4 and return to constant or less bluer status in Phase 5. \label{cc_overall}}
\end{figure}
\clearpage

\begin{figure}
\epsscale{.80}
\centering
\plotone{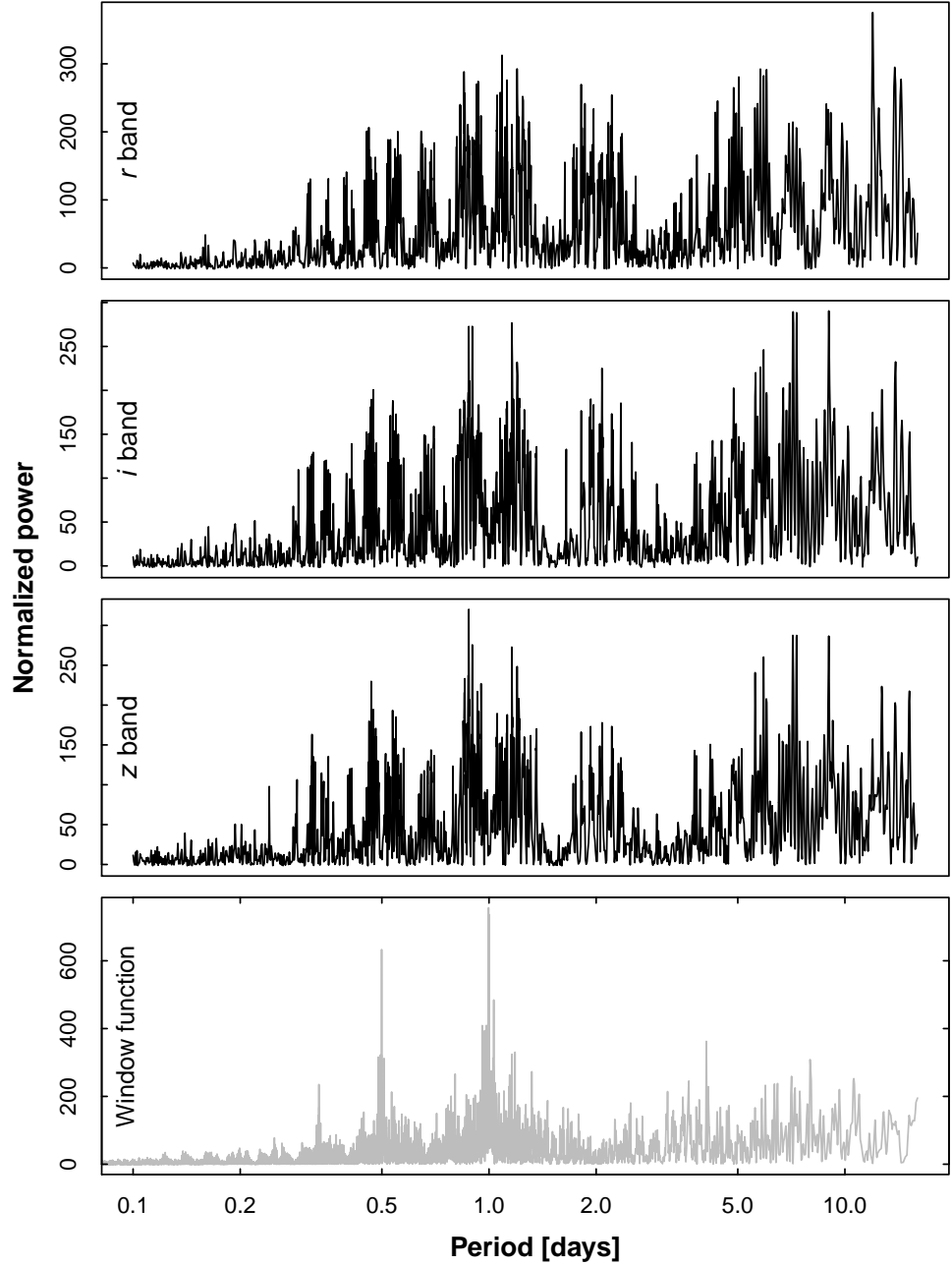}
\caption{\label{periodograms} Lomb-Scargle periodogram of the data at each band. The bottom panel shows the periodogram of the window function.}
\end{figure}

\begin{deluxetable}{lcccc}
\tabletypesize{\scriptsize}
\centering
\tablecolumns{5}
\tablewidth{0pc}
\tablecaption{Best-fit solutions in each band.\label{solutions}}
\tablehead{
& \colhead{$\mathrm{1^{st}}$ per. [d]} & \colhead{$\mathrm{2^{nd}}$ per. [d]} & \colhead{$\log\mathcal{L}$} & \colhead{$\chi^2$}}
\startdata
\tableline
\multicolumn{5}{l}{\textbf{\textit{r} band}}\\
\tableline
No periodicity &  & & 747.57 & 32.9 \\ 
One periodicity &	&  11.97 & 431.91 & 17.3 \\
Two periodicities &	11.99 & 7.15 & 293.08 & 10.7 \\
\tableline
\multicolumn{5}{l}{\textbf{\textit{i} band}}\\
\tableline
No periodicity & &  &  434.45 & 16.37 \\ 
One periodicity & & 1.16 & 218.44 & 10.25 \\
Two periodicities & 9.03 & 0.95 & 124.57 & 6.68 \\
\tableline
\multicolumn{5}{l}{\textbf{\textit{z} band}}\\
\tableline
No periodicity & &  &  241.41 & 9.73 \\ 
One periodicity &   & 0.88 &  42.60 & 5.76 \\ 
Two periodicities & 9.02 & 0.94 & -46.48 & 4.00 \\
\enddata
\end{deluxetable}

\begin{figure}[ht]
\centering
\rotatebox{90}{
\begin{minipage}[c]{\textwidth}
\hspace{0.3cm}
\includegraphics[width=0.3\textwidth]{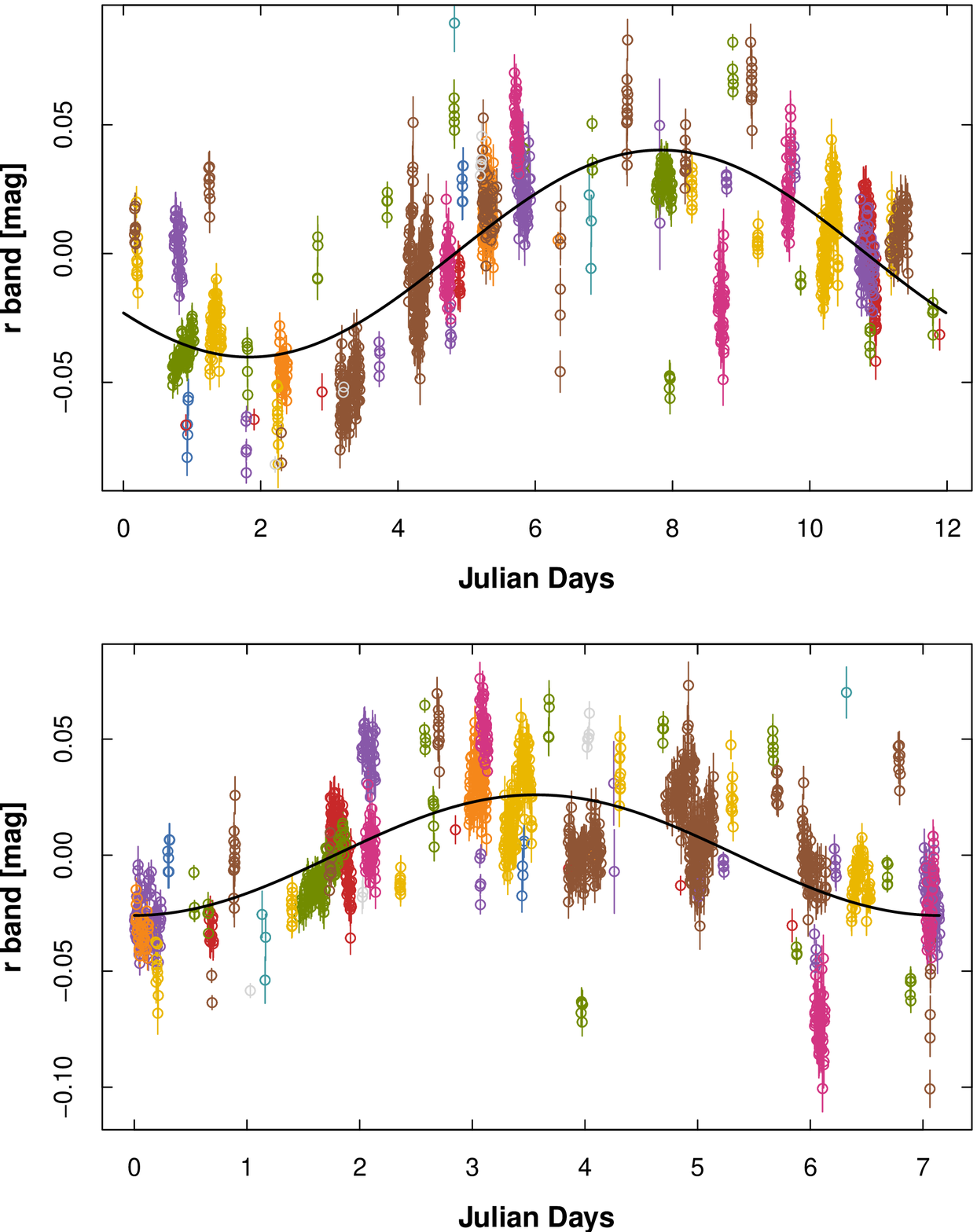}
\includegraphics[width=0.3\textwidth]{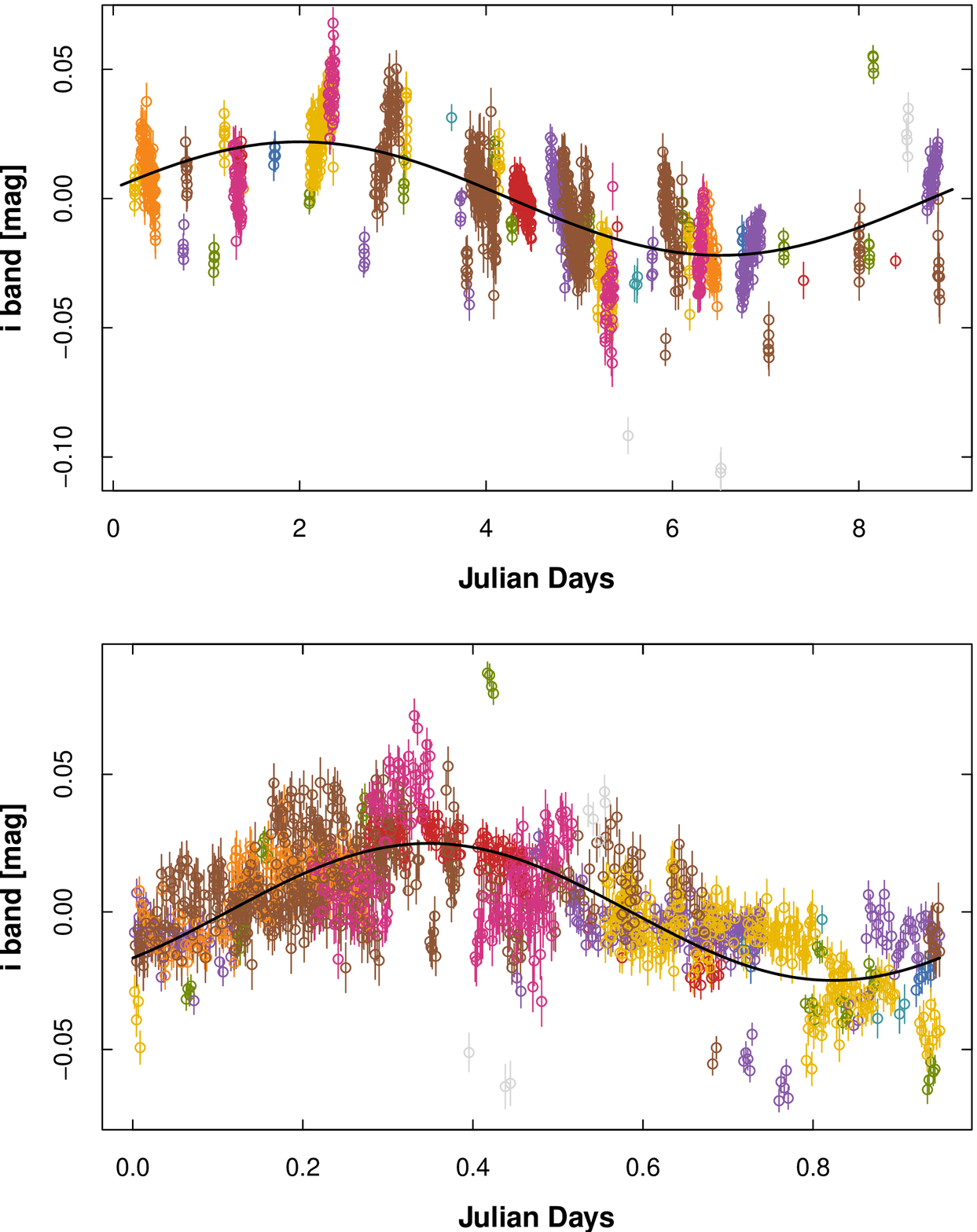}
\includegraphics[width=0.3\textwidth]{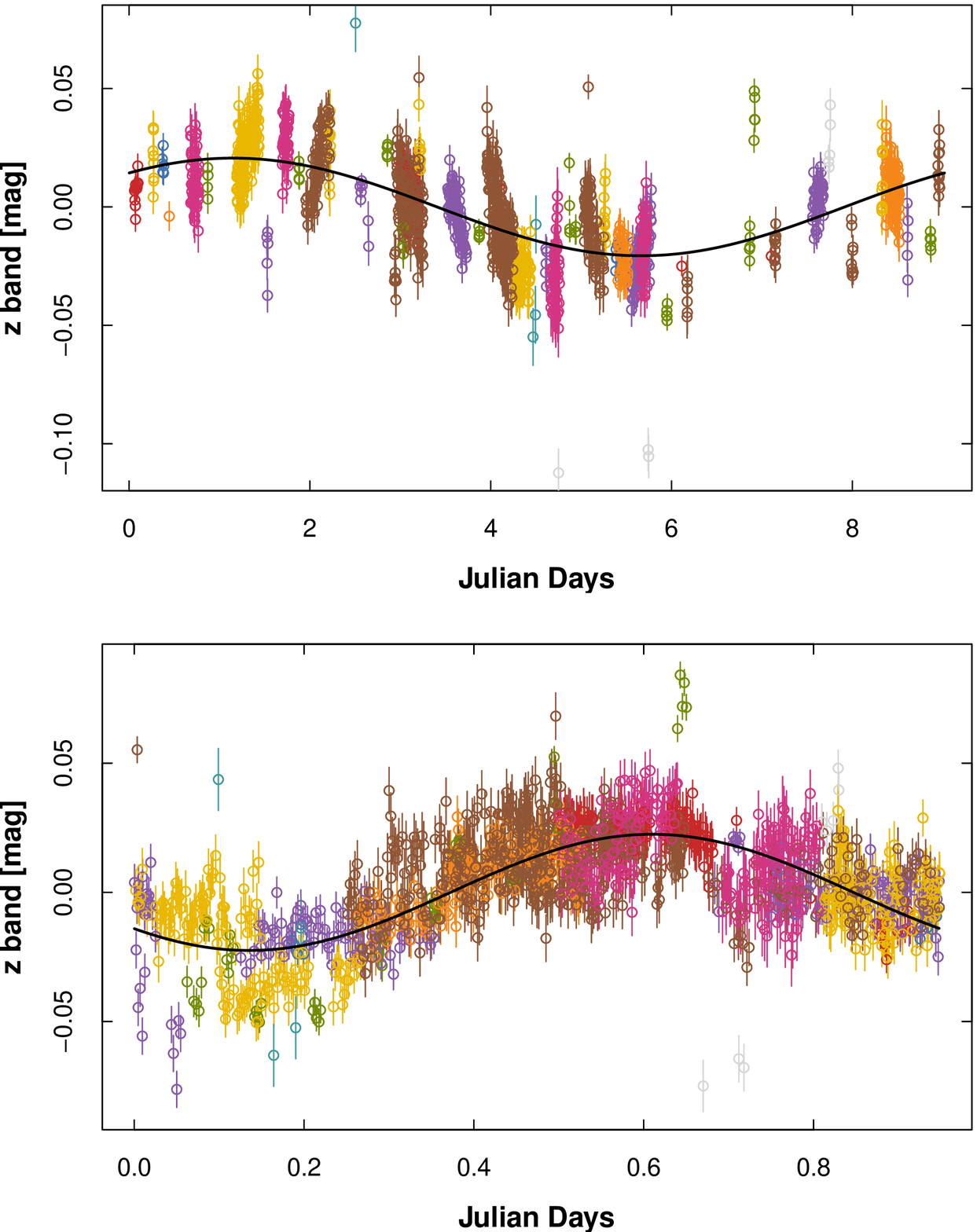}
\hspace{0.3cm}
\caption{\label{bestfits} Phased best-fitting models for the data obtained in $r$, $i$ and $z$ bands. Each observing run that is shifted with a floating offset is marked with a different color. Note that alternative models are also possible (Table \ref{solutions}).}
\end{minipage}}
\end{figure}

\begin{figure}
\epsscale{.80}
\plotone{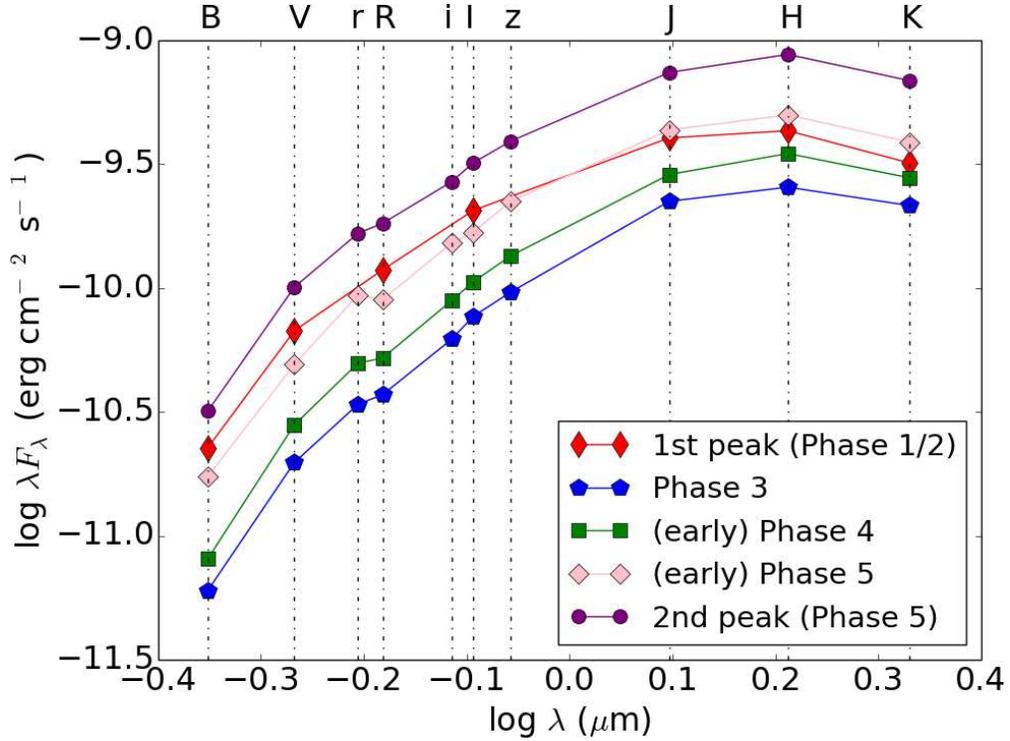}
\caption{Multi-epoch SEDs of HBC 722 after outburst. $B$, $V$, $R$, and $I$ bands are from \citet{semkov2012b,semkov2014}. $r$, $i$ and $z$ bands are taken from this work. $J$, $H$ and $K_{S}$ data are taken from \citet{kospal2011}, \citet{anto2013} and \citet{sung2013}. All data except $r$, $i$ and $z$ are converted from Vega to AB magnitudes system using \citet{br2007}. \label{sed}}
\end{figure}

\clearpage

\begin{figure}
\epsscale{.80}
\plotone{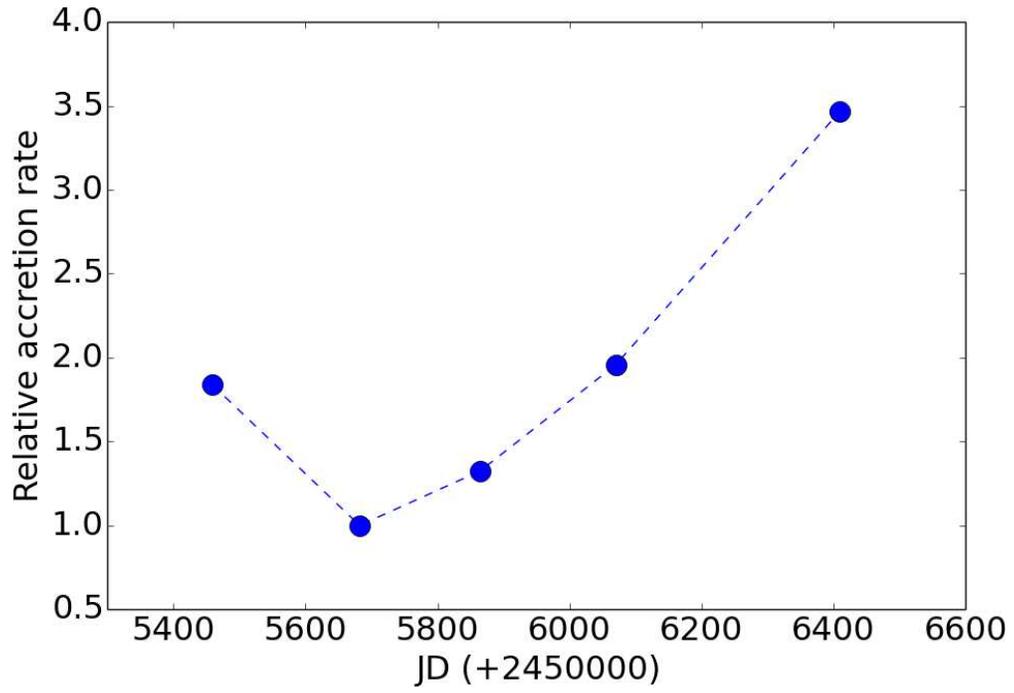}
\caption{Relative accretion rate in each phase. We normalize derived accretion rate values to the minimum brightness epoch after outburst (2011 April) to compare the change of accretion rate in the post outburst phase. \label{acc}}
\end{figure}

\clearpage
\begin{deluxetable}{cccccc}
\tabletypesize{\scriptsize}
\tablecolumns{6}
\tablewidth{0pc}
\tablecaption{Result of day scale variability of HBC 722.\label{var_day}}
\tablehead{
\colhead{Month} & \colhead{Filter} & \colhead{Duration [Day]} & \colhead{C$_{r}$} & \colhead{C$_{i}$} & \colhead{C$_{z}$}}
\startdata
2011-07 &  $r$,$i$,$z$ &6 & 2.54 & 2.73 &1.83\\
2011-08 &  $r$,$i$,$z$ &13 & 4.86& 6.63& 4.69\\
2011-11 &  $r$,$i$,$z$ &9 & 2.44 & 2.42& 1.59\\
2012-06 &  $r$,$i$,$z$ &7 & 4.97&  4.34& 3.29\\
2012-09 &  $r$,$i$,$z$ &11 & 2.68 & 1.98 & 1.39 \\
2013-05 &  $r$,$i$,$z$ &4 &  3.39 & 3.93 & 2.83\\
\enddata
\tablecomments{C values represent relative brightness derivation of HBC 722 to that of comparison star. It provides statistical confidences of variability of the source. C value above 1.64, 1.96 and 2.57 infer confidence level of the variability in 90\%, 95\% and 99\% respectively.}
\end{deluxetable}

\clearpage

\begin{deluxetable}{ccccccc}
\tabletypesize{\scriptsize}
\tablecolumns{7}
\tablewidth{0pc}
\tablecaption{Result of intra-day variability (IDV) of HBC 722.\label{var_intraday}}
\tablehead{
\colhead{Date (UT)} & \colhead{Filter} & \colhead{Number of frames} & \colhead{Observing time [hour]} & \colhead{C$_{r}$} & \colhead{C$_{i}$} & \colhead{C$_{z}$}}
\startdata
2011-04-26 &  $r$,$i$,$z$ &24 & 0.5 &1.10 & 0.80 &1.19\\
2011-07-05 &  $r$,$i$,$z$ &311 & 4.5 &2.11 &2.04 &1.48\\
2011-07-11 &  $r$,$i$,$z$ &47 & 0.9 &1.16 & 1.47&1.04\\
2011-08-24 &  $r$ &1125 & 7.5 &1.68 & \nodata & \nodata\\
2011-08-31 &  $r$ & 791 & 5.7 &1.43 & \nodata & \nodata \\
2011-11-02 & $r$,$i$,$z$ &211 & 5.3 &1.50  &1.08  & 0.82\\
2011-11-04 &  $r$,$i$,$z$ &116 & 2.9 &1.71  &1.24  & 0.98\\
2011-11-09 &  $r$,$i$,$z$ &206 & 4.9 &1.50 &1.32  & 1.06\\
2012-05-27 &  $r$,$i$,$z$ &128 & 2.7 &0.90 & 1.24 & 0.82\\
2012-05-30 &  $r$,$i$,$z$ &300 & 3.8 &1.45  &0.96  &0.90\\
2012-06-28 & $r$,$i$,$z$ &399 & 6.2 &1.42 &0.88 &0.96 \\
2012-07-01 & $r$,$i$,$z$ &182 & 4.0 &1.45 & 1.05 &0.88 \\
2012-07-02 &  $r$,$i$,$z$ &32 & 0.5 &1.06 & 1.24 &0.94\\
2012-09-01 &  $r$,$i$,$z$ &328 & 6.2 &1.07 &1.22 &1.18 \\
2012-09-02 & $r$,$i$,$z$ &492 & 7.1 &0.97 &2.29 &1.01 \\
2012-09-03 & $r$,$i$,$z$ &219 & 5.2 &1.03 &1.39 &0.87 \\
2012-09-06 &  $r$,$i$,$z$ &23 & 0.5 &1.98 &1.89 &1.13\\
2012-09-07 &  $r$,$i$,$z$ &31 & 0.5 &1.19 &0.98 &0.82\\
2012-09-09 & $r$,$i$,$z$ &177 & 6.3 &1.24 &1.52 &0.92 \\
2012-11-26 &  $r$,$i$,$z$ &53 & 0.5 &0.93 &1.06 &1.00\\
2013-04-30 &  $r$,$i$,$z$ &162 & 2.2 &0.99 &1.10 &0.86\\
2013-05-01 &  $r$,$i$,$z$ &144 & 1.7 &1.24 &2.20 &0.98\\
2013-05-04 &  $r$,$i$,$z$ &143 & 2.2 &0.99 &1.03 &0.80\\
2013-05-05 &  $r$,$i$,$z$ &150 & 1.8 &1.68 &1.70 &0.76\\
\enddata
\tablecomments{C values for the confidence level follow the same rule in Table \ref{var_day}.}
\end{deluxetable}

\clearpage

\begin{figure}
\epsscale{.80}
\plotone{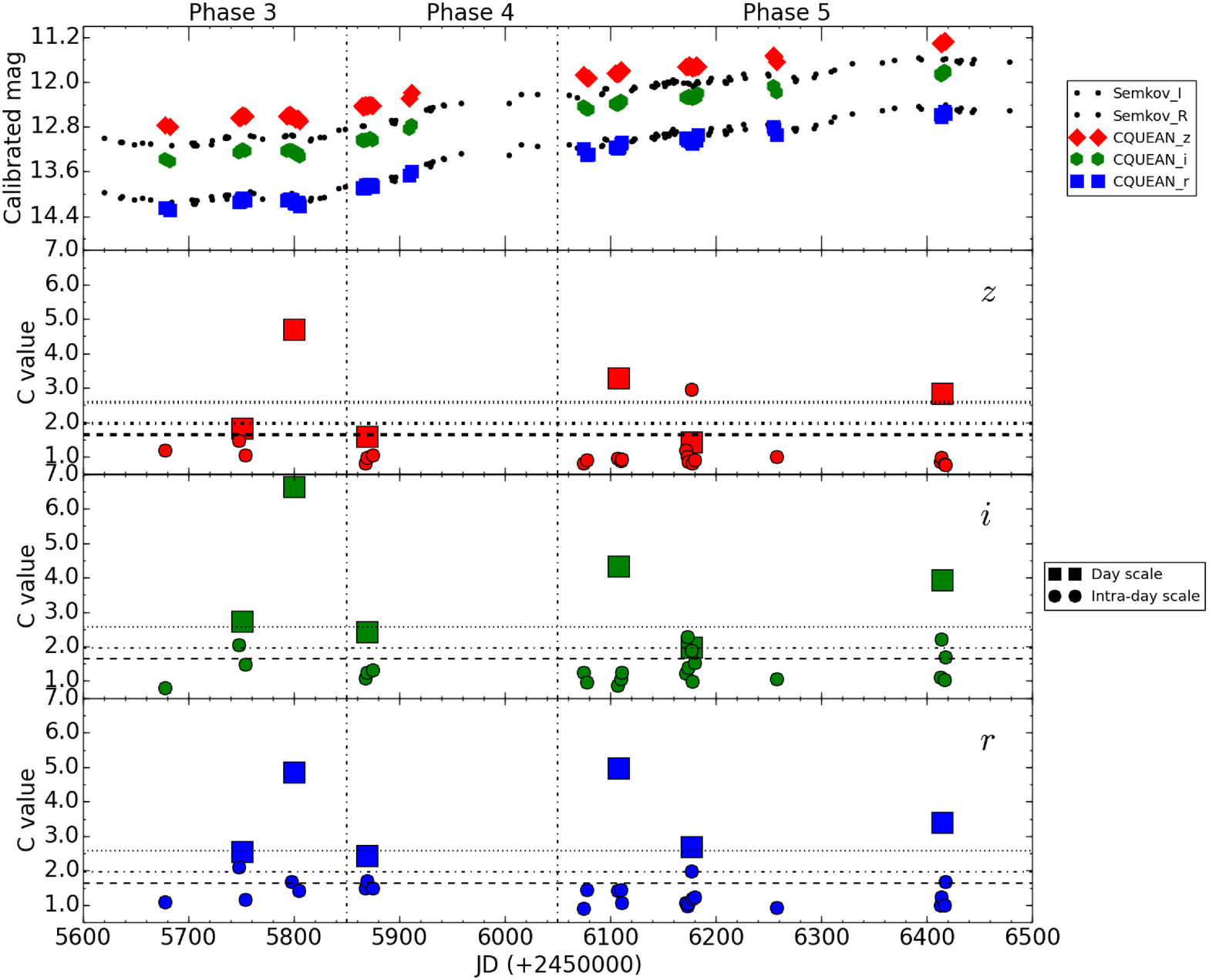}
\caption{Overall light curves and plots of C values. The top plot shows light curves from 2011 April to 2013 May as a reference for long term brightness variation. The upper (red), middle (green) and lower (blue) curves depict $z$, $i$ and $r$ bands and $I$ and $R$ bands references are also described (black). From the second to the fourth plots show C values distribution for $z$, $i$ and $r$ bands, respectively. Large squares and small circles indicate day scale from Table \ref{var_day} and intra-day scale from Table \ref{var_intraday}, respectively. \label{C}}
\end{figure}

\clearpage


\begin{thebibliography}{}
\bibitem[Audard et al.(2014)]{audard2014} Audard, M., \'{A}brah\'{a}m, P., Dunham, M., M. et al. 2014, review chapter in Protostars and Planets VI, University of Arizona Press (2014), eds. H. Beuther, R. Klessen, C. Dullemond, Th. Henning, 1401, 3368, in press
\bibitem[Antoniucci et al.(2013)]{anto2013} Antoniucci, S., Arkharov, A., Klimanov, S., et al. 2013, ATel, 5023,1
\bibitem[Bastien et al.(2011)]{bastien2011} Bastien, F. A.,  Stassun, K. G., Weintraub, D. A. 2011, \aj, 142, 141
\bibitem[Bell \& Lin(1994)]{bl1994} Bell, K. R., \& Lin, D. N. C. 1994, \apj, 427, 987
\bibitem[Bertin \& Arnouts(1996)]{ba1996} Bertin, E., \& Arnouts, S. 1996, \aaps, 117, 393
\bibitem[Blanton \& Roweis(2007)]{br2007} Blanton, M. R., \& Roweis, S. 2007, \aj, 133, 734
\bibitem[Chambers(1983)]{chambers83} Chambers, J. 1983, Graphical Methods for Data Analysis (Wadsworth)
\bibitem[Clarke et al.(2005)]{clarke2005} Clarke, C., Lodato, G., Melnikov, S. Y., Ibrahimov., M., A. 2005, MNRAS, 361, 942
\bibitem[Cohen \& Kuhi(1979)]{ck1979} Cohen, M., \& Kuhi, L. V. 1979, ApJS, 41, 743
\bibitem[Fukugita et al.(1996)]{fj1996} Fukugita, M., Ichikawa, T., Gunn, J. E., et al. 1996, \aj, 111, 1748
\bibitem[Goodrich(1987)]{goodrich1987} Goodrich, R. W. 1987, \pasp, 99, 116
\bibitem[Green et al.(2013)]{green2013} Green, J. D., Robertson, P., Baek, G., et al. 2013, \apj, 764, 22
\bibitem[Gupta et al.(2008)]{gupta2008} Gupta, A. C., Cha, S.-M., Lee, S., et al. 2008, \aj, 136, 2359
\bibitem[Hartmann \& Kenyon(1996)]{hk1996} Hartmann, L., \& Kenyon, S. J. 1996, \araa, 34, 207
\bibitem[Hartmann(2008)]{h2008} Hartmann, L. 2008, Accretion Processes in Star Formation Second Edition, (Cambridge astrophysics series, 47; Cambridge: Cambridge Univ. Press)
\bibitem[Herbig(2003)]{herbig2003} Herbig, G. H., Petrov, P. P., Duemmler, R. 2003, \apj, 595, 384
\bibitem[Johnstone et al.(2013)]{johnstone2013} Johnstone, D., Hendricks, B., Herczeg, G., Bruderer, S. 2013, \apj, 765,133
\bibitem[Kim et al.(2011)]{kim2011} Kim, E., Park, W.-K., Jeong, H., et al. 2011, JKAS, 44, 115
\bibitem[Kenyon et al.(2000)]{kenyon2000} Kenyon, S. J., Kolotilov, E. A., Ibragimov, M. A., Mattei, J., A. 2000, \apj, 531, 1028
\bibitem[K\'{o}sp\'{a}l et al.(2011)]{kospal2011} K\'{o}sp\'{a}l, \'{A}., \'{A}brah\'{a}m, P., Acosta-Pulido, J. A. et al. 2011, \aap, 527, 133
\bibitem[Lim et al.(2013)]{lim2013} Lim, J., Chang, S., Pak, S., et al., 2013, JKAS, 46, 161
\bibitem[Lee et al.(2011)]{lee2011} Lee, J.-E., Kang, W., Lee, S.-G., et al. 2011, JKAS, 44, 67
\bibitem[Laugalys et al.(2006)]{laugalys2011} Laugalys, V., Strai\v{z}ys, V., Vrba, F. J., et al., 2006, Baltic Astronomy, 15, 483
\bibitem[Miller et al.(2011)]{miller2011} Miller, A. A., Hillenbrand, L. A., Covey, K. R., et al. 2011, \apj, 730, 80
\bibitem[Jang \& Miller(1997)]{jm1997} Jang, M., Miller, H. R. 1997, \aj, 114, 565
\bibitem[Park et al.(2012)]{park2012} Park, W.-K., Pak, S., Im, M., et al. 2012, \pasp, 124, 839
\bibitem[Powell et al.(2012)]{powell2012} Powell, S. L., Irwin, M., Bouvier, J., Clarke, C., J. 2012, MNRAS, 426, 3315
\bibitem[Romero et al.(1999)]{romero1999} Romero, G. E., Cellone, S. A., Combi, J. A. 1999, \aaps, 135, 477
\bibitem[Semkov et al.(2010)]{semkov2010} Semkov, E. H., Peneva, S. P., Munari, U., et al. 2010, \aap, 523, 3
\bibitem[Semkov et al.(2012a)]{semkov2012a} Semkov, E. H., Peneva, S. P., Munari, U., et al. 2012a, yCat, 354, 29043
\bibitem[Semkov et al.(2012b)]{semkov2012b} Semkov, E. H., Peneva, S. P., Munari, U., et al. 2012b, \aap, 542, 43
\bibitem[Semkov et al.(2014)]{semkov2014} Semkov, E. H., Peneva, S. P., Ibryamov, S., I., Dimitrov, D., P. 2014, BlgAJ, 20, 59
\bibitem[Shu et al.(1994)]{shu1994} Shu, F., Najita, J., Ostriker, E., et al. 1994, \apj, 429, 781
\bibitem[Siwak et al.(2013)]{siwak2013} Siwak, M., Rucinski, S. M. Matthews, J. M., et al. MNRAS, 432, 194
\bibitem[Skrutskie et al.(2006)]{2mass}Skrutskie, M. F., Cutri, R. M., Stiening, R., et al. 2006, \aj, 131, 1163
\bibitem[Smith et al.(2002)]{smith2002} Smith, J., A., Tucker, D. L., Kent, S., et al. 2002, \aj, 123, 2121.
\bibitem[Sung et al.(2013)]{sung2013} Sung, H.-I., Park, W.-K., Yang, Y., et al. 2013, JKAS, 46, 253.
\bibitem[Weintraub et al.(1991)]{wein1991} Weintraub, D. A., Sandell, G., Duncan, W. D. 1991, \apj, 382, 270
\bibitem[Wright et al.(2010)]{wise} Wright, E., L., Eisenhardt, P., R., M., Mainzer, A., K., et al. 2010, \aj, 140, 1868
\bibitem[Zechmeister \& K{\"u}rster(2009)]{zechmeister2009} Zechmeister, M., \& {K{\"u}rster}, M. 2009, \aap, 496, 577
\end{thebibliography}
\end{document}